\documentclass{article}
\usepackage[utf8]{inputenc}

\title{Synthetic Data Generation for Economists \vspace{-1ex}}
\author{Allison Koenecke \thanks{\texttt{koenecke@stanford.edu}, Stanford Institute for Computational \& Mathematical Engineering} 
\and Hal Varian \thanks{Google Economics Team}}
\date{Presented at the American Economic Association (AEA) Annual Meeting \\January 2020 \vspace{-7ex}}

\usepackage{graphicx}
\usepackage{geometry}
\usepackage{footmisc}
\usepackage{setspace}
\usepackage{titling}

\begin{document}
\newgeometry{top=0.9in,bottom=0.9in}
\begin{spacing}{1.2}
\maketitle

\section{Motivation}

As more tech companies engage in rigorous economic analyses, we are confronted with a data problem: in-house papers cannot be replicated due to use of sensitive, proprietary, or private data.  Readers are left to assume that the obscured true data (e.g., internal Google information) indeed produced the results given, or they must seek out comparable public-facing data (e.g., Google Trends) that yield similar results \cite{superbowl}.  One way to ameliorate this reproducibility issue is to have researchers release synthetic datasets based on their true data; this allows external parties to replicate an internal researcher's methodology.  In this brief overview, we explore synthetic data generation at a high level for economic analyses.

One analogy explaining synthetic data generation involves OpenAI's GPT-2 model for text generation \cite{gpt2}.  The model takes as input several starter lines of text, and outputs additional text based on the prompt; when the user gives the model just one sentence, she receives a full story back.  In economic applications, we instead feed numerical data (such as time series values) into a generative model.  The model then produces similar but new data that preserve certain aspects of the structure, such as covariances across attributes.  In this way, we construct a plausible extension of the seed -- but not a true extension that would reveal private information.

The ideal outline for future publications using sensitive, proprietary, or private data looks roughly as follows:
\begin{enumerate}
    \item Describe the true data.
    \item Describe a synthetic data generating model that produces a synthetic dataset.
    \item Run the same analyses on both true and synthetic data.
    \item Confirm that results on both datasets are comparable.
    \item Publicly release the synthetic dataset.
\end{enumerate}

The above framework both allows an external agent to reproduce the internal researcher's methodology, and allows a separate internal agent to reproduce the synthetic data via their access to the true data.
With this in mind, we turn to describing a variety of generative models that may be useful for economists.

\section{Generative Models for Synthetic Data}

We begin with the caveat that there is no single catch-all method to best generate synthetic data: one must consider the ensuing analyses before choosing the data generating method.  For example, if only the first and second moments need to be captured, generating synthetic data will not be nearly as computationally intensive as an analysis requiring higher moments.  Further, if the data must be scrubbed to enforce differential privacy when performing summary statistics, then additional precautions should be taken when generating synthetic data.

A relatively basic but comprehensive method for data generation is the Synthetic Data Vault (SDV) \cite{sdv}.  This work uses the multivariate Gaussian Copula when calculating covariances across input columns. Then, the distributions and covariances are sampled to form synthetic data.  As proof of concept, five relational datasets were synthetically generated and used by freelance data scientists to develop predictive models.  The researchers found no significant difference between the results produced using the synthetic versus true data.

While the SDV allows for a basic perturbation of model parameters to create noisy versions of the data, it does not explicitly address more concrete data privacy concerns.  There is, however, a vast literature in the machine learning community regarding differential privacy\footnote{We provide the formal definition for $\epsilon$-differential privacy here.  A randomized algorithm $\mathcal{A}: \mathcal{D} \rightarrow \mathcal{R}$ with domain $\mathcal{D}$ and range $\mathcal{R}$ is ($\epsilon$, $\delta$)-differentially private if for any two adjacent training datasets d, d' $\subseteq$ D, which differ by at most one training point, and any subset of outputs $\mathcal{S} \subseteq \mathcal{R}$, it satisfies that: 
$\textrm{Pr}[\mathcal{A}(d) \in \mathcal{S}] \leq e^{\epsilon} \textrm{Pr}[\mathcal{A}(d')  \in \mathcal{S}] + \delta$, where $\epsilon$ is the privacy budget and $\delta$ is a failure rate \cite{diff-pri}. } 
with generative models; a Gaussian mechanism is often used to perturb the input with noise sampled from a normal distribution based on the sensitivity \cite{dwork}.  Within a machine learning framework, ``privacy loss'' can be quantified at each iteration of training, which ends when the loss reaches a pre-defined privacy budget.  This training is often performed on auto-encoders\footnote{These are models used in unsupervised learning, which use dimensionality reduction to hone in on the most important data features, and are particularly useful for high-dimensional input data.  The superficially similar Variational Auto-Encoder \cite{vae} encourages its latent variables to be distributed as a Gaussian distribution, enabling use as a generative model.}, which learn the latent structure of an input (using the ``encoder''), and then reconstruct the input (using the ``decoder''); these two parts of the network are generally trained together to minimize reconstruction loss.  In particular, the auto-encoder technique DP-SYN has been shown experimentally to out-perform other methods of privacy-preserving synthetic data generation \cite{sweeney}.  Further, it is important for differential privacy to defend against model inversion and Generative Adversarial Network\footnote{We discuss GANs at length in the following section.} (GAN)-based attacks; these types of attacks are able to reconstruct training data despite the training process being differentially private \cite{fed-attack}, even when using federated learning (as popularized by Google \cite{fed-learn}).  The harm from these attacks can be mitigated by using auto-encoder-based generative model techniques such as DP-AuGM (which requires the user to include public data seeds to generate new data based on the private data training), and DP-VaeGM (which generates an infinite amount of data based on Gaussian noise, but is less stable) \cite{diff-pri}.  The examples cited here are applied to classification tasks; there additionally exist algorithms for differentially private regression \cite{linreg-diff1, linreg-diff2} and association rule mining \cite{rules-1, rules-2} that can be applied to synthetic data.

Next, we discuss the current literature with regard to robustness.  Economists are familiar with using torture tests to generate a worst-case model and finding the minimal perturbation necessary to break a result (e.g., reversing a sign in some variable of interest).  Robustness checks on the data rather than the model can be done to confirm that perturbing the true data (when generating synthetic data) is not problematic.  In the deep learning world, similar analyses are a cornerstone of adversarial machine learning research; much of it follows the canonical 2013 paper showing that data perturbations can result in misclassification of images \cite{img}.  More recent work has studied neural network robustness necessary for resistance against adversarial attacks \cite{attack}.  Robustness research is not limited to deep learning; work has also been done using robust regressions with low-rank approximations to defend against what has been termed ``training data poisoning'' \cite{linreg}.  In this vein, economists can further confirm the transferability of results obtained from their sensitive, proprietary, or private data.

\section{Generative Adversarial Networks}

Further to the above generative methods described, there has been much excitement surrounding Generative Adversarial Networks (GANs) \cite{gans} in generating synthetic data.  While common applications include generation of images and text, there has recently been increasing interest in the economic space.  We first explore what GANs are, and then briefly survey the current literature as it relates to generating synthetic data for economists.

GANs generate data using two components: a generator (which creates candidates for data distributions) and a discriminator (which evaluates the candidates)\footnote{Usually, the generator and discriminator are various kinds of neural networks.  But, we can also imagine a simple GAN wherein both generator and discriminator are simply parametric models.  The generator is trained to optimize for increasing the error rate of the discriminator.}.
If the discriminator is able to correctly classify the generator's candidates as synthetic data (as opposed to true data), then the generator has failed.  The generator will try to act adversarially to fool the discriminator into not being able to differentiate between the true and synthetic data (i.e., minimizing the maximum distance between these datasets). Wasserstein GANs, in particular, use the Wasserstein distance\footnote{Intuitively, the Wasserstein metric can be thought of in relation to the optimal transport problem, wherein the goal is to transport one distribution of mass to a different distribution (of the same mass) on the same space.  We can think of this as shifting a pile of sand from one distribution to another; which grains of sand should be moved so that the cost of transporting sand is minimized?}
between probability distributions on a given metric space \cite{wgan}.

In the economic literature, GANs are a useful way to redesign standard Monte Carlo studies. As one example, Wasserstein GANs have been used to generate realistic synthetic data by Athey et al. \cite{athey}, which were then used to compare different estimators for average treatment effects.  In addition to using a systematic simulation study design, a new estimator has been proposed based on GAN discriminators, and was found to be more efficient than a standard indirect inference estimator \cite{kaji}.  This is broadly useful in structural estimation where researchers aim to learn about policy effects arising from economic models that often have intractable likelihoods.  Lastly, it is of note that privacy-preserving methods can also be extended to GAN-generated synthetic data, in applications ranging from image generation \cite{image-gen} to clinical studies \cite{gan-priv}.  

While we only cover a few generative models in this overview, it is worth noting that there are many other types of deep generative models worth exploring, including: Variational Auto-Encoders \cite{vae}, Autoregressive Models (MADE \cite{made}, Pixel RNN \cite{pixel-rnn}, Pixel CNN++ \cite{pixel-cnn}, WaveNet \cite{wavenet}), Normalizing Flow Models (RealNVP \cite{real-nvp}, Glow \cite{glow}), and Energy-Based Models \cite{ebm}.  Going forward, we hope to see more synthetic data generation methods in the economic literature.


\end{spacing}

\bibliographystyle{plain}
{\footnotesize
\bibliography{references}}

\begin{thebibliography}{10}

\bibitem{sweeney}
Nazmiye~Ceren Abay, Yan Zhou, Murat Kantarcioglu, Bhavani Thuraisingham, and
  Latanya Sweeney.
\newblock Privacy preserving synthetic data release using deep learning.
\newblock In Michele Berlingerio, Francesco Bonchi, Thomas G{\"a}rtner, Neil
  Hurley, and Georgiana Ifrim, editors, {\em Machine Learning and Knowledge
  Discovery in Databases}, pages 510--526, Cham, 2019. Springer International
  Publishing.

\bibitem{wgan}
Martin Arjovsky, Soumith Chintala, and Léon Bottou.
\newblock Wasserstein gan, 2017.

\bibitem{athey}
Susan Athey, Guido Imbens, Jonas Metzger, and Evan Munro.
\newblock Using wasserstein generative adversarial networks for the design of
  monte carlo simulations, 2019.

\bibitem{gan-priv}
Brett~K. Beaulieu-Jones, Zhiwei~Steven Wu, Chris Williams, Ran Lee, Sanjeev~P.
  Bhavnani, James~Brian Byrd, and Casey~S. Greene.
\newblock Privacy-preserving generative deep neural networks support clinical
  data sharing.
\newblock {\em bioRxiv}, 2018.

\bibitem{linreg-diff1}
Kamalika Chaudhuri and Claire Monteleoni.
\newblock Privacy-preserving logistic regression.
\newblock In D.~Koller, D.~Schuurmans, Y.~Bengio, and L.~Bottou, editors, {\em
  Advances in Neural Information Processing Systems 21}, pages 289--296. Curran
  Associates, Inc., 2009.

\bibitem{diff-pri}
Qingrong Chen, Chong Xiang, Minhui Xue, Bo~Li, Nikita Borisov, Dali Kaarfar,
  and Haojin Zhu.
\newblock Differentially private data generative models, 2018.

\bibitem{real-nvp}
Laurent Dinh, Jascha Sohl-Dickstein, and Samy Bengio.
\newblock Density estimation using real nvp, 2016.

\bibitem{dwork}
Cynthia Dwork, Frank McSherry, Kobbi Nissim, and Adam Smith.
\newblock Calibrating noise to sensitivity in private data analysis.
\newblock In {\em Theory of Cryptography}, volume Vol. 3876, pages 265--284, 01
  2006.

\bibitem{made}
Mathieu Germain, Karol Gregor, Iain Murray, and Hugo Larochelle.
\newblock Made: Masked autoencoder for distribution estimation.
\newblock {\em Proceedings of the 32nd International Conference on Machine
  Learning, JMLR W\&CP 37:881-889, 2015}, 2015.

\bibitem{gans}
Ian~J. Goodfellow, Jean Pouget-Abadie, Mehdi Mirza, Bing Xu, David
  Warde-Farley, Sherjil Ozair, Aaron Courville, and Yoshua Bengio.
\newblock Generative adversarial networks, 2014.

\bibitem{fed-attack}
Briland Hitaj, Giuseppe Ateniese, and Fernando Perez-Cruz.
\newblock Deep models under the gan: Information leakage from collaborative
  deep learning, 2017.

\bibitem{kaji}
Tetsuya Kaji, Elena Manresa, and Guillaume Pouliot.
\newblock Deep inference: Artificial intelligence for structural estimation.
\newblock {\em Barcelona GSE}, 2018.

\bibitem{glow}
Diederik~P. Kingma and Prafulla Dhariwal.
\newblock Glow: Generative flow with invertible 1x1 convolutions, 2018.

\bibitem{vae}
Diederik~P Kingma and Max Welling.
\newblock Auto-encoding variational bayes, 2013.

\bibitem{ebm}
Yann LeCun, Sumit Chopra, Raia Hadsell, Fu~Jie Huang, and et~al.
\newblock A tutorial on energy-based learning.
\newblock In {\em Predicting Structured Data}. MIT Press, 2006.

\bibitem{rules-1}
Ninghui Li, Wahbeh Qardaji, Dong Su, and Jianneng Cao.
\newblock Privbasis: Frequent itemset mining with differential privacy.
\newblock {\em Proceedings of the VLDB Endowment (PVLDB), Vol. 5, No. 11, pp.
  1340-1351 (2012)}, 2012.

\bibitem{linreg}
Chang Liu, Bo~Li, Yevgeniy Vorobeychik, and Alina Oprea.
\newblock Robust linear regression against training data poisoning.
\newblock In {\em Proceedings of the 10th ACM Workshop on Artificial
  Intelligence and Security}, AISec '17, pages 91--102, New York, NY, USA,
  2017. ACM.

\bibitem{attack}
Aleksander Madry, Aleksandar Makelov, Ludwig Schmidt, Dimitris Tsipras, and
  Adrian Vladu.
\newblock Towards deep learning models resistant to adversarial attacks, 2017.

\bibitem{fed-learn}
H.~Brendan McMahan, Eider Moore, Daniel Ramage, Seth Hampson, and
  Blaise~Agüera y~Arcas.
\newblock Communication-efficient learning of deep networks from decentralized
  data.
\newblock {\em Proceedings of the 20 th International Conference on Artificial
  Intelligence and Statistics (AISTATS) 2017. JMLR: W\&CP volume 54}, 2016.

\bibitem{sdv}
N.~{Patki}, R.~{Wedge}, and K.~{Veeramachaneni}.
\newblock The synthetic data vault.
\newblock In {\em 2016 IEEE International Conference on Data Science and
  Advanced Analytics (DSAA)}, pages 399--410, Oct 2016.

\bibitem{gpt2}
Alec Radford, Jeff Wu, Rewon Child, David Luan, Dario Amodei, and Ilya
  Sutskever.
\newblock Language models are unsupervised multitask learners.
\newblock {\em OpenAI blog}, 2019.

\bibitem{pixel-cnn}
Tim Salimans, Andrej Karpathy, Xi~Chen, and Diederik~P. Kingma.
\newblock Pixelcnn++: Improving the pixelcnn with discretized logistic mixture
  likelihood and other modifications, 2017.

\bibitem{superbowl}
Seth Stephens-Davidowitz, Hal Varian, and Michael~D. Smith.
\newblock {Super returns to Super Bowl ads?}
\newblock {\em Quantitative Marketing and Economics (QME)}, 15(1):1--28, March
  2017.

\bibitem{img}
Christian Szegedy, Wojciech Zaremba, Ilya Sutskever, Joan Bruna, Dumitru Erhan,
  Ian Goodfellow, and Rob Fergus.
\newblock Intriguing properties of neural networks, 2013.

\bibitem{wavenet}
Aaron van~den Oord, Sander Dieleman, Heiga Zen, Karen Simonyan, Oriol Vinyals,
  Alex Graves, Nal Kalchbrenner, Andrew Senior, and Koray Kavukcuoglu.
\newblock Wavenet: A generative model for raw audio, 2016.

\bibitem{pixel-rnn}
Aaron van~den Oord, Nal Kalchbrenner, and Koray Kavukcuoglu.
\newblock Pixel recurrent neural networks, 2016.

\bibitem{image-gen}
Liyang Xie, Kaixiang Lin, Shu Wang, Fei Wang, and Jiayu Zhou.
\newblock Differentially private generative adversarial network, 2018.

\bibitem{rules-2}
C.~Zeng, J.~F. Naughton, and J.~Y. Cai.
\newblock {{O}n {D}ifferentially {P}rivate {F}requent {I}temset {M}ining}.
\newblock {\em VLDB J}, 6(1):25--36, Nov 2012.

\bibitem{linreg-diff2}
Jun Zhang, Zhenjie Zhang, Xiaokui Xiao, Yin Yang, and Marianne Winslett.
\newblock Functional mechanism: Regression analysis under differential privacy.
\newblock {\em Proceedings of the VLDB Endowment (PVLDB), Vol. 5, No. 11, pp.
  1364-1375 (2012)}, 2012.

\end{thebibliography}

\end{document}